\documentclass[twocolumn,showpacs,amsfonts,amsmath,amssymb,floatfix,superscriptaddress]{revtex4}
\usepackage{url}
\usepackage{bm}		
\usepackage{graphicx}	

\begin{document}


\title[The interior structure of rotating black holes]{Towards a general description of the interior structure of rotating black holes}

\author{Andrew J S Hamilton}
\email{Andrew.Hamilton@colorado.edu}	
\affiliation{JILA, Box 440, U. Colorado, Boulder, CO 80309, USA}
\affiliation{Dept.\ Astrophysical \& Planetary Sciences,
U. Colorado, Boulder, CO 80309, USA}

\newcommand{\simpropto}{\raisebox{-0.7ex}[1.5ex][0ex]{
		\begin{array}[b]{@{}c@{\;}} \propto \\
		[-1.8ex] \sim \end{array}}}

\newcommand{\dd}{d}
\newcommand{\ddsq}{\dd^2\mkern-1.5mu}
\newcommand{\ddd}{\dd^3\mkern-1.5mu}
\newcommand{\dddd}{\dd^4\mkern-1.5mu}
\newcommand{\DD}{D}
\newcommand{\ee}{e}
\newcommand{\im}{i}
\newcommand{\Ei}{{\rm Ei}}
\newcommand{\perpperp}{\perp\!\!\perp}
\newcommand{\ppartial}{\partial^2\mkern-1mu}
\newcommand{\nn}{\nonumber\\}

\newcommand{\diag}{{\rm diag}}
\newcommand{\jel}{\text{\sl j}}
\newcommand{\Lz}{L}
\newcommand{\Msun}{{\rm M}_\odot}
\newcommand{\uel}{\text{\sl u}}
\newcommand{\vel}{\text{\sl v}}
\newcommand{\inn}{{\rm in}}
\newcommand{\out}{{\rm ou}}
\newcommand{\sep}{{\rm sep}}

\newcommand{\bg}{\bm{g}}
\newcommand{\bp}{\bm{p}}
\newcommand{\bv}{\bm{v}}
\newcommand{\bx}{\bm{x}}
\newcommand{\bgamma}{\bm{\gamma}}

\newcommand{\Deltax}{\Delta_x}
\newcommand{\Deltay}{\Delta_y}
\newcommand{\KCarter}{{\cal K}}
\newcommand{\Mass}{{\cal M}}
\newcommand{\mbh}{m_\bullet}
\newcommand{\Mbh}{M_\bullet}
\newcommand{\Mbhdot}{\dot{M}_\bullet}
\newcommand{\NUT}{{\cal N}}
\newcommand{\Qelec}{Q}
\newcommand{\Qelecbh}{\Qelec_\bullet}
\newcommand{\Qmag}{{\cal Q}}
\newcommand{\Qmagbh}{\Qmag_\bullet}
\newcommand{\rhosep}{\rho_{\rm Kerr}}
\newcommand{\rhox}{\rho_x}
\newcommand{\rhoy}{\rho_y}
\newcommand{\unit}[1]{\, {\rm{#1}}}

\hyphenpenalty=3000

\begin{abstract}
The purpose of this paper is to present a number of proposals
about the interior structure of a rotating black hole
that is accreting slowly,
but in an arbitrary time- and space-dependent fashion.
The proposals could potentially be tested with numerical simulations.
Outgoing and ingoing particles free-falling in the parent Kerr geometry
become highly focused along the principal outgoing and ingoing null directions
as they approach the inner horizon,
triggering the mass inflation instability.
The original arguments of
Barrab\'es, Israel \& Poisson (1990)
regarding inflation in rotating black holes
are reviewed, and shown to be based on Raychauduri's equation
applied along the outgoing and ingoing null directions.
It is argued that gravitational waves should behave in the geometric optics
limit, and consequently that the spacetime should be almost shear-free.
A full set of shear-free equations is derived.
A specific line-element is proposed,
which is argued should provide a satisfactory approximation
during early inflation.
Finally,
it is argued that 
super-Planckian collisions between outgoing and ingoing particles
will lead to entropy production,
bringing inflation to an end, and precipitating collapse.
\end{abstract}


\date{\today}

\maketitle

\section{Introduction}

What happens if you fall inside an astronomical black hole?
Singularity theorems are quite unspecific
\cite{Senovilla:1997},
stating only that, subject to
a trapped surface condition,
an energy condition,
and
a causality condition,
a geodesic is incomplete.
Singularity theorems do not prescribe the nature of the incompleteness,
nor do they assert that any matter actually moves along the incomplete geodesic,
and they certainly do not imply that an infalling observer must fall
to a singularity.

A major advance was made when Poisson \& Israel
\cite{Poisson:1989zz,Poisson:1990eh}
discovered the mass inflation instability at the inner horizon
of a charged, spherical (Reissner-Nordstr\"om) black hole.
The inflationary instability is the nonlinear realization
of the infinite blueshift at the inner horizon first pointed out by Penrose
\cite{Penrose:1968}.
Many subsequent analytic and numerical investigations
confirmed the instability (see \cite{Hamilton:2008zz}
for a review and references).
Chan, Chan \& Mann
\cite{Chan:1994rs}
extended the Poisson-Israel argument to an
azimuthally symmetrically accreting
rotating black hole in 1+2 dimensions.

Barrab\`es, Israel \& Poisson
\cite{Barrabes:1990},
hereafter BIP,
soon generalized
Poisson \& Israel's argument to the case of a rotating black hole.
The BIP argument is reviewed and critiqued below, \S\ref{BIP}.
Further efforts to explore inflation in rotating black holes
were made by Ori
\cite{Ori:1992,Ori:2001pc}.

In a recent series of papers
\cite{Hamilton:2010a,Hamilton:2010b,Hamilton:2010c},
the author presented some explicit solutions for the interior
structure of a rotating black hole that
slowly accretes a collisionless fluid.
The solutions followed the self-consistent nonlinear evolution
of the spacetime through inflation
to collapse down to an exponentially tiny scale,
where rotation reasserted itself.
However, the solutions were limited to the conformally separable special case,
where the accretion flow incident on the inner horizon was uniform.

The purpose of the present paper is to
set forward arguments about how inflation should develop
in a rotating black hole that is accreting slowly,
but in an arbitrary time- and space-dependent fashion.
The arguments are in places qualitative rather than rigorous,
so should be construed as tentative.
Specifically,
the argument that the spacetime is likely to be almost shear-free,
\S\ref{gravitationalwaves},
the possibility that the line-element~(\ref{lineelement})
may provide an adequate approximation,
and the argument that inflation will be terminated
by super-Planckian collisions between outgoing and ingoing particles,
\S\ref{bhpa},
are proposals to be tested.
The proposals could be tested by numerical simulations.
Whether true or false, I hope that simulators will find the ideas herein
a useful guide as to what might be expected.

At first sight,
it might seem that the problem of the interior structure
of a rotating black hole accreting in an arbitrary
time- and space-dependent fashion
would be too complex to be analytically tractable.
However, the problem simplifies
if the black hole is accreting slowly,
and therefore has a geometry well described by Kerr
away from its inner horizon.
Aside from rare (but interesting!)\ events of high accretion,
such as when a black hole first forms,
or when two black holes merge,
real astronomical black holes accrete slowly
for most of their lives.

A slowly accreting black hole has
a number of simplifying features.
First,
as shown by
\cite{Hamilton:2010b},
freely-falling particles focus along the principal
outgoing and ingoing null directions of the Kerr geometry
as they approach the inner horizon,
regardless of the initial orbital parameters of the particles.
Moreover, for small accretion rates,
outgoing and ingoing particles are already
hyper-relativistic relative to each other when inflation ignites,
so massive as well as massless, and charged as well as uncharged
particles, all focus along the principal null directions
as they approach the inner horizon.
Thus the situation originally envisaged by BIP
of outgoing and ingoing null streams crossing near the inner horizon
is in fact realized in the real case.

The second simplifying feature is that
during inflation,
although the streaming energy-momentum and the Weyl curvature
grow exponentially huge,
the underlying geometry remains scarcely changed.
In a sense, nothing is happening,
despite the exponentially growing curvature.
The reason for this is that the proper time
experienced by an infaller during inflation is tiny,
so that volume elements have little time to become distorted
despite the enormous tidal forces,
a fact first pointed out by \cite{Ori:1991}.

Eventually however the spacetime does respond to the huge acceleration,
by collapsing.

The metric signature in this paper is
${-}{+}{+}{+}$.

\section{The BIP argument}
\label{BIP}

It is helpful to start by summarizing BIP's argument.
For generality,
I will recast BIP's
\cite{Barrabes:1990}
argument into continuum form.
As will be seen,
BIP's argument
is essentially founded on the Raychaudhuri equation applied
along two null directions
(whereas Penrose's original singularity theorem
\cite{Penrose:1965}
invoked the Raychaudhuri equation
along a single null direction).
In \S\ref{critique},
I point out issues left open by BIP's argument.

\subsection{Argument}

BIP
start by positing two null shells that cross each other
near the inner horizon of a rotating black hole.
As noted in the Introduction,
freely-falling particles do in fact focus along the principal
outgoing and ingoing null directions of the Kerr geometry
as they approach the inner horizon,
regardless of the initial orbital parameters or mass of the particles
\cite{Hamilton:2010b}.
Thus the situation of two special null directions envisaged by
BIP does in fact occur in the real case,
at least for a slowly accreting black hole.

The principal null directions of the Kerr geometry are geodesic.
More generally,
as the geometry departs from Kerr
thanks to the inflationary back-reaction,
the two special null directions posited by BIP can be taken
to be the geodesic continuation of the principal null directions.
Choose a Newman-Penrose double-null tetrad frame
$\{ \bgamma_v , \bgamma_u , \bgamma_+ , \bgamma_- \}$
such that the outgoing and ingoing tetrad axes
$\bgamma_v$ and $\bgamma_u$
point along the two special outgoing and ingoing null geodesic directions,
while the spinor axes
$\bgamma_+$ and $\bgamma_-$
(which are complex conjugates of each other)
span the two-dimensional spatial plane orthogonal to the null directions.

In what follows,
it is convenient to follow a convention in which
early latin indices $a$, $b$, ... run over spinor indices $+$, $-$,
late latin indices $z$, $y$, ... run over null indices $v$, $u$,
and mid latin indices $k$, $l$, ... run over all four indices.

The condition that the special null directions be geodesic
means that the tetrad-frame momenta $p^a$ orthogonal to the geodesics satisfy
$\DD p^a / \DD \lambda = 0$,
where $\DD$ denotes the covariant derivative,
and $\lambda$ is an affine parameter.
This in turn implies the vanishing of 4
of the 24 distinct tetrad-frame connections
$\Gamma_{klm}$
(recall that the tetrad-frame connections $\Gamma_{klm}$,
being generators of Lorentz transformations,
are antisymmetric in their first two indices,
$\Gamma_{klm} = - \Gamma_{lkm}$),
namely
\begin{equation}
\label{nosheara}
  \Gamma_{+vv} = \Gamma_{-vv} = \Gamma_{+uu} = \Gamma_{-uu} = 0
  \ .
\end{equation}
These conditions use up four of the six gauge degrees of freedom
of Lorentz transformations of the tetrad.
The two remaining tetrad degrees of freedom correspond to
a Lorentz boost in the $vu$ plane,
and
a spatial rotation in the $+-$ plane.

The extrinsic curvatures along the two null directions $v$ and $u$
are defined in the usual way to be the two $2 \times 2$ matrices
$\Gamma_{avb}$
and
$\Gamma_{aub}$.
The extrinsic curvatures are tetrad-frame tensors with respect to
the two unfixed gauge degrees of freedom of the tetrad,
a fact proven in Appendix~\ref{proofextrinsic}.
Being tetrad-frame quantities,
all tetrad-frame connections $\Gamma_{klm}$
are automatically coordinate gauge-invariant.
The standard definition of the tetrad-frame Riemann tensor
$R_{klmn}$,
coupled with the conditions~(\ref{nosheara}),
implies the generalized Raychaudhuri equations
for the extrinsic curvatures
along each of the null directions
$z = v, u$
(no implicit summation over the two lowered indices $z$),
\begin{equation}
\label{Raychaudhuriv}
  \DD_z \Gamma_{a z b}
  +
  \Gamma_{azc} \Gamma^c{}_{z b}
  +
  R_{zazb}
  =
  0
  \ .
\end{equation}
The four components of the extrinsic curvature
$\Gamma_{azb}$
along each null direction
are commonly decomposed into
a trace part, the expansion $\vartheta_z$,
an antisymmetric part, the twist $\varpi_z$,
and a traceless symmetric part, the complex shear $\sigma_z$:
\begin{subequations}
\label{Sachs}
\begin{align}
\label{Sachsv}
  \Gamma_{+v-}
  =
  \Gamma^\ast_{-v+}
  \equiv
  \vartheta_v + \im \varpi_v
  &\ , \quad
  \Gamma_{+v+}
  =
  \Gamma^\ast_{-v-}
  \equiv
  \sigma_v
  \ ,
\\
\label{Sachsu}
  \Gamma_{-u+}
  =
  \Gamma^\ast_{+u-}
  \equiv
  \vartheta_u + \im \varpi_u
  &\ , \quad
  \Gamma_{-u-}
  =
  \Gamma^\ast_{+u+}
  \equiv
  \sigma_u
  \ .
\end{align}
\end{subequations}
In terms of the expansion, twist, and shear,
the generalized Raychaudhuri equations~(\ref{Raychaudhuriv})
along the outgoing null direction $v$ are
\begin{subequations}
\label{RaychaudhuriSachs}
\begin{align}
\label{Raychaudhuriexpansion}
  ( \DD_v + \vartheta_v ) \vartheta_v
  -
  \varpi_v^2
  +
  \sigma_v \sigma^\ast_v
  +
  4\pi
  T_{vv}
  &=
  0
  \ ,
\\
\label{Raychaudhuritwist}
  ( \DD_v + 2 \vartheta_v ) \varpi_v
  &=
  0
  \ ,
\\
\label{Raychaudhurishear}
  ( \DD_v + 2 \vartheta_v ) \sigma_v
  +
  C_{v+v+}
  &=
  0
  \ ,
\end{align}
\end{subequations}
where $C_{klmn}$ is the Weyl tensor,
the traceless part of the Riemann tensor $R_{klmn}$.
A similar set of equations holds for the ingoing direction $u$,
with $C_{v+v+}$ replaced by $C_{u-u-}$
in equation~(\ref{Raychaudhurishear}) for the shear.
The covariant derivative $\DD_v$ in equations~(\ref{RaychaudhuriSachs})
can be interpreted as acting either on rank-1 vectors
$\vartheta_v$, $\varpi_v$, and $\sigma_v$,
or on their rank-3 tensor antecedents, equations~(\ref{Sachsv});
the result is identical, given conditions~(\ref{nosheara}).
In the Raychaudhuri equation~(\ref{Raychaudhuriexpansion})
for the expansion,
Einstein's equations have been invoked to replace the Ricci tensor
$R_{vv}$
by the energy-momentum
$8\pi T_{vv}$.
The energy-momentum $T_{vv}$
that goes into the equation for the outgoing expansion
is the ingoing energy flux:
if the energy-momentum is characterized
as a sum over streams of particles with tetrad-frame
number currents $n^k$ and momenta $p^k$,
then the ingoing energy flux $T_{vv}$ is
\begin{equation}
  T_{vv} = T^{uu} = \sum_{\rm streams} n^u p^u
  \ .
\end{equation}

In the BIP scenario of two null shells crossing,
the Raychaudhuri equations~(\ref{Raychaudhuriexpansion})
can be integrated immediately over the crossing shells,
leading to the result that the expansions
$\vartheta_v$
and
$\vartheta_u$
in the outgoing and ingoing directions
are changed respectively by the integrals
$4\pi \int T_{vv} \, \dd v$
and
$4\pi \int T_{uu} \, \dd u$
over the delta-function of flux in the opposing direction.

BIP pointed out that the
product $\vartheta_v \vartheta_u$
of the null expansions is a tetrad-frame scalar
(with respect to the two unfixed tetrad degrees of freedom),
as well as a coordinate scalar,
and they proposed that this
product
would define an effective scalar mass function,
as it does in spherically symmetric black holes.
In fact for a (non-inflating) Kerr black hole
of mass $\Mbh$ and specific angular momentum $a$, the
product
of the expansions is
(in standard Boyer-Lindquist spheroidal coordinates)
\begin{equation}
  \vartheta_v
  \vartheta_u
  =
  {r^2 ( r^2 {+} a^2 ) \over 2 ( r^2 + a^2 \cos^2\!\theta )^3}
  \left( {2 \Mbh r \over r^2 {+} a^2} - 1 \right)
  \ .
\end{equation}
BIP argued that the closer to the inner horizon
that the shells passed through each other,
the more blue-shifted the energy-momentum in the opposite shell would appear,
and therefore the greater the increase of mass.


\subsection{Critique}
\label{critique}

As BIP acknowledge,
their analysis falls short of a complete description
of inflation in rotating  black holes.

At least initially,
and for small accretion rates,
the assumption that energy-momentum is focused along two
special null directions appears robust,
since outgoing and ingoing particles necessarily focus along
the principal null directions of the Kerr geometry
as they approach the inner horizon
\cite{Hamilton:2010b}.
It seems highly likely that geodesics will remain focused,
because
the Raychaudhuri equation~(\ref{Raychaudhuriexpansion})
indicates that the effect of energy-momentum focused
along the outgoing and ingoing directions is to cause further focusing
(make the expansion more negative).

BIP's assumption of two thin shells
allows the Raychaudhuri equations~(\ref{Raychaudhuriexpansion})
to be integrated immediately,
but neglects the effects of twist and shear.
Are twist and shear in fact unimportant?
In the conformally separable solutions of
\cite{Hamilton:2010a,Hamilton:2010b,Hamilton:2010c},
inflation was followed by collapse to an exponentially tiny scale,
where rotation reasserted itself.
The finite angular momentum of a null geodesic bundle is encoded in its twist.
The Raychaudhuri equation~(\ref{Raychaudhuritwist})
shows that twist is amplified by collapse,
and the Raychaudhuri equation~(\ref{Raychaudhuriexpansion})
shows that the effect of twist is to slow collapse.
The Raychaudhuri equation~(\ref{Raychaudhurishear})
shows similarly that shear is amplified by collapse.
Notwithstanding the argument in \S\ref{gravitationalwaves}
that shear is probably small,
shear will probably also become important following collapse.
Whereas twist slows collapse, shear speeds it.

This suggests that twist and shear are probably unimportant prior to collapse,
but are likely to become important following collapse.
The subsequent evolution of the spacetime
could be complicated, possibly with BKL-like behavior
\cite{Belinski:1970,Belinski:1971,Belinski:1982,Berger:2002st,Ashtekar:2011ck}.

A potentially more serious defect is that
the BIP analysis does not take into account
the possible back-reaction of the inflating energy-momenta on the horizon.
In the solutions of
\cite{Hamilton:2010a,Hamilton:2010b,Hamilton:2010c},
inflation is followed by collapse,
a fate inaccessible to BIP's analysis.


Ideally,
it would be nice to have an explicit, general solution
that self-consistently follows the evolution of the spacetime
and energy-momentum through inflation to whatever happens then.
The intent of this paper is to make some proposals
about what that solution might look like.

\section{The role of gravitational waves}
\label{gravitationalwaves}

In this section it is argued that shear is likely to be small.

BIP, following
\cite{Poisson:1990eh},
suggested that the outgoing energy-momentum needed for inflation to occur
would be provided by a Price \cite{Price:1972} tail of gravitational radiation
generated by the collapse of the black hole.
On the other hand
\cite{Hamilton:2010b}
argued that in real astronomical black holes,
direct accretion of baryons and dark matter
from outside the outer horizon provides a continual
source of both outgoing and ingoing streams incident on the inner horizon,
that probably soon overwhelm the Price tails generated by collapse.
Direct accretion of matter
can provide not only ingoing but also outgoing streams.
Although a stream falling through the outer horizon
is necessarily ingoing, it will become outgoing at the inner horizon
if its angular momentum is large enough
in the same direction as the angular momentum of the black hole.

This raises the question of how important gravitational waves are
in shaping the structure of spacetime during inflation and thereafter?

One place where gravitational waves play an essential role
is in radiating away the ``hair'' that a black hole acquires
when it accretes anisotropically.
The loss of hair by gravitational radiation is what makes
it reasonable to treat a slowly accreting black hole
as being approximated accurately by the Kerr geometry
down to near its inner horizon.

However,
from the perspective of what is happening in the inflationary zone
near the inner horizon,
gravitational waves should behave
in the high-frequency, geometric-optics limit,
as if they were any other kind of (massless) particle,
as both BIP and \cite{Hamilton:2010b} argued.
Averaged over regions much larger than a wavelength,
the gravitational wave contribution to the Einstein tensor
can be taken over to the right hand side of the Einstein equations
and reinterpreted as ``real'' energy-momentum.

Mathematically,
the proposition that
gravitational waves behave in the geometric optics limit
is equivalent to the assertion that
the high-frequency spin~$\pm 2$ components of the Weyl tensor
vanish when averaged over the rapidly oscillating field,
\begin{equation}
\label{Weylaverage}
  \langle C_{z+z+} \rangle
  =
  \langle C_{z-z-} \rangle
  =
  0
  \ .
\end{equation}
It then follows from integrating
the Raychaudhuri equation~(\ref{Raychaudhurishear})
that the average high-frequency shear also vanishes,
given that the shear is initially zero,
as it is in the Kerr geometry,
\begin{equation}
\label{noshearaverage}
  \langle \sigma_z \rangle
  =
  0
  \ .
\end{equation}
On the other hand,
the absolute value squared of the high-frequency shear does not vanish
on averaging, and indeed could possibly be substantial,
\begin{equation}
\label{shearsquared}
  \langle \sigma_z \sigma^\ast_z \rangle
  \neq
  0
  \ .
\end{equation}
This averaged mean square shear~(\ref{shearsquared})
makes a contribution to the Raychaudhuri equation~(\ref{Raychaudhuriexpansion})
for the expansion,
behaving effectively as if it were
equivalent to energy-momentum $4\pi T_{zz}$.

The analogous situation in electromagnetism is
that electromagnetic waves behave as photons,
in which the high-frequency components of the electric and magnetic fields
vanish on average,
but their squares do not vanish, and carry energy-momentum.

What about shear at long wavelengths, comparable to the size of the black hole?
For slow accretion,
the geometry away from the inner horizon will be close to Kerr,
which is shear-free.
Thus even long-wavelength shear is likely to be small.

Of course, just as photons can scatter, so also gravitons can scatter.
However, I argue in \S\ref{bhpa}
that scattering processes are likely to become important
only when center-of-mass collision energies exceed the Planck scale.

\section{Shear-free equations}
\label{completeequations}

If shear is negligible,
as argued in the previous section \S\ref{gravitationalwaves},
then a more complete set of equations can be derived,
still without any explicit line-element.

\subsection{Equations}
\label{equations}

As remarked in \S\ref{BIP},
the 8 tetrad-frame connections embodied by the extrinsic curvatures
along the two null directions form tetrad-frame tensors
with respect to the two unfixed tetrad gauge degrees of freedom.
A further 8 tetrad-frame connections are tetrad-frame tensors,
namely the extrinsic curvatures
$\Gamma_{yaz}$
along the angular tetrad directions $a = + , -$.
The extrinsic curvature along each angular direction
can be decomposed into a trace part, the expansion $\vartheta_a$,
and an antisymmetric part, the twist $\varpi_a$,
analogously to equations~(\ref{Sachs}),
\begin{subequations}
\begin{align}
  {\textstyle \frac{1}{2}} (
  \Gamma_{+vu}
  +
  \Gamma_{+uv}
  )
  \equiv
  \vartheta_+
  &\ , \quad
  {\textstyle \frac{1}{2}} (
  \Gamma_{+vu}
  -
  \Gamma_{+uv}
  )
  \equiv
  \im \varpi_+
\\
  {\textstyle \frac{1}{2}} (
  \Gamma_{-uv}
  +
  \Gamma_{-vu}
  )
  \equiv
  \vartheta_-
  &\ , \quad
  {\textstyle \frac{1}{2}} (
  \Gamma_{-uv}
  -
  \Gamma_{-vu}
  )
  \equiv
  \im \varpi_-
  \ .
\end{align}
\end{subequations}
Note that
$\vartheta_- = \vartheta_+^\ast$
and
$\varpi_- = \varpi_+^\ast$.
The angular shears $\sigma_a$ vanish identically,
equations~(\ref{nosheara}).

The 16 tetrad-frame connections
comprising the extrinsic curvatures along each of the 4 tetrad axes
satisfy the 16 generalized Raychaudhuri equations
\begin{equation}
\label{Raychaudhuriz}
  \DD_z \Gamma_{a z k}
  +
  \Gamma_{azb} \Gamma^b{}_{z k}
  +
  R_{zazk}
  =
  0
  \ ,
\end{equation}
where, as before,
early latin indices $a$, $b$ run over spinor indices $+$, $-$,
the late latin index $z$ runs over null indices $v$, $u$,
and the mid latin index $k$ runs over all four indices.
Only the paired index $b$ with one raised and one lowered is summed over;
the pair of lowered $z$ indices is not summed over.
The 16 equations~(\ref{Raychaudhuriz})
include 4 whose terms all vanish identically,
namely the equations for the angular shear $\Gamma_{azz}$.
Equations~(\ref{Raychaudhuriz})
are valid without any assumption about the spacetime:
they hold given any arbitrary pair of null geodesic directions,
and a tetrad aligned with them.

Further progress can be made if,
as argued in \S\ref{gravitationalwaves},
the shear along the two null directions can be neglected,
equation~(\ref{noshearaverage}),
or equivalently
\begin{equation}
\label{noshearz}
  \Gamma_{+v+}
  =
  \Gamma_{-v-}
  =
  \Gamma_{+u+}
  =
  \Gamma_{-u-}
  =
  0
  \ .
\end{equation}
Under the condition~(\ref{noshearz}) of vanishing shear,
the extrinsic curvatures along each of the 4 tetrad directions
satisfy a further 16 generalized Raychaudhuri equations,
\begin{equation}
\label{Raychaudhuria}
  \DD_a \Gamma_{z a k}
  +
  \Gamma_{zay} \Gamma^y{}_{a k}
  +
  R_{azak}
  =
  0
  \ .
\end{equation}
The two sets of Raychaudhuri equations~(\ref{Raychaudhuriz})
and (\ref{Raychaudhuria})
comprise two distinct sets of 8 non-vanishing equations for
the expansions and twists
along each of the tetrad directions,
a total of 16 non-vanishing equations altogether.

Combining the
Raychaudhuri equations~(\ref{Raychaudhuriz}) and (\ref{Raychaudhuria}),
and invoking Einstein's equations to replace the Ricci tensor
by energy-momenta,
yields 8 Einstein equations,
\begin{subequations}
\label{Einstein}
\begin{align}
\label{Einsteinzz}
  \DD_z \Gamma^a{}_{z a}
  +
  \Gamma^a{}_{zc} \Gamma^c{}_{z a}
  +
  8\pi
  T_{zz}
  &=
  0
  \ ,
\\
\label{Einsteinza}
  \DD_z \Gamma_{azy}
  +
  \DD_a \Gamma_{azb}
  +
  2 \Gamma_{azb} \Gamma_{azy}
  +
  8\pi T_{za}
  &=
  0
  \ ,
\\
\label{Einsteinaa}
  \DD_a \Gamma^z{}_{a z}
  +
  \Gamma^z{}_{ay} \Gamma^y{}_{a z}
  +
  8\pi
  T_{aa}
  &=
  0
  \ .
\end{align}
\end{subequations}
In equation~(\ref{Einsteinza}),
$y$ is the null index opposite to $z$,
and
$b$ is the spinor index opposite to $a$.

The Raychaudhuri equations~(\ref{Raychaudhuriz}) and (\ref{Raychaudhuria})
combine in a different way
to yield an expression for the 4 spin~$\pm 1$ components
$C_{zazy} = C_{azab}$
of the Weyl tensor,
\begin{equation}
  \DD_z \Gamma_{azy}
  -
  \DD_a \Gamma_{azb}
  +
  2 C_{zazy}
  =
  0
  \ ,
\end{equation}
where again
$y$ is the null index opposite to $z$,
and
$b$ is the spinor index opposite to $a$.
Altogether the 16 non-vanishing generalized Raychaudhuri equations
comprise 8 Einstein equations, 4 Weyl equations,
and 4 twist equations.

The above equations account for all but 8 of the equations
relating the Riemann tensor to
derivatives of the tetrad connections.
The remaining 8 equations involve the scalar
(with respect to the two unfixed degrees of freedom)
components of the tetrad-frame Riemann tensor.
Four of the equations involve only the expansion and twist connections.
Of these, one consitutes an equation for
a combination of the spin~$0$ polar (real)
Weyl component $C_{vuvu}$ and the Ricci scalar $R$,
and another constitutes an equation for
the spin~$0$ axial (imaginary) Weyl component $C_{vu+-}$:
\begin{subequations}
\label{Weylscalar}
\begin{align}
  &
  R_{v+u-}
  +
  R_{v-u+}
  =
  - \,
  C_{vuvu}
  -
  {\textstyle \frac{1}{6}}
  R
\\
\nonumber
  &=
  \DD^z
  \vartheta_z
  +
  \DD^a
  \vartheta_a
  + \,
  \vartheta^z \vartheta_z
  +
  \vartheta^a \vartheta_a
  +
  \varpi^z \varpi_z
  +
  \varpi^a \varpi_a
  \ ,
\\
  &
  C_{vu+-}
  =
  R_{vu+-}
  =
  \im
  \left(
  \DD^z \varpi_z + \DD^a \varpi_a
  \right)
  \ .
\end{align}
\end{subequations}
In equations~(\ref{Weylscalar}),
radial and angular vectors
(super/sub-scripted $z$ and $a$ respectively)
are to be interpreted as
lying in their respective 2-dimensional tangent spaces
(thus for example $\vartheta_z = \{ \vartheta_v , \vartheta_u , 0 , 0 \}$),
and the covariant derivatives act on the corresponding 2-dimensional space.
The two other equations involving only expansions and twists are:
\begin{subequations}
\begin{align}
  \DD_v \vartheta_u - \DD_u \vartheta_v
  &=
  \im \left[
  ( \DD_+ + 2 \vartheta_+ ) \varpi_-
  -
  ( \DD_- + 2 \vartheta_- ) \varpi_+ 
  \right]
  \ ,
\\
  \DD_+ \vartheta_- - \DD_- \vartheta_+
  &=
  \im \left[
  ( \DD_v + 2 \vartheta_v ) \varpi_u
  -
  ( \DD_u + 2 \vartheta_u ) \varpi_v
  \right]
  \ .
\end{align}
\end{subequations}

The remaining 4 Riemann equations
include 2 that determine the remaining scalar components
$R_{vuvu}$ and $R_{+-+-}$
of the tetrad-frame Riemann tensor.
These two components
cannot be expressed solely in terms of expansions and twists
and their covariant derivatives.
Rather, they depend on two further gauge-invariant quantities
that correspond physically to the Lorentz boost between
the affinely-parameterized outgoing and ingoing null directions,
and the spatial rotation angle between
the affinely-parameterized angular directions.
An explicit example will be seen in \S\ref{modellineelement}.

If the energy-momenta $T_{kl}$
are arranged to satisfy covariant conservation of energy-momentum,
as they should,
and if the 8 Einstein equations~(\ref{Einstein}) are satisfied,
then the Einstein tensor will automatically satisfy covariant conservation,
as it should,
\begin{equation}
\label{DG}
  \DD^k G_{kl}
  =
  0
  \ .
\end{equation}
The conservation equations~(\ref{DG})
provide 2 evolution equations for $G_{vu}$
(along each of the null directions),
and 2 constraint equations for $G_{+-}$
(along each of the spinor directions),
that will be satisfied automatically.

%

\subsection{Schematic solution of the equations}
\label{scheme}

The equations
of the previous subsection, \S\ref{equations},
were derived
without the benefit of any explicit line-element
or choice of coordinates,
the only assumption being that shear vanishes,
as argued in \S\ref{gravitationalwaves}.
It is helpful to outline how
these equations could in principle be solved.

Suppose that the energy-momenta $T_{kl}$ are given.
As argued in \S\ref{bhpa},
outgoing and ingoing streams should stream through each other collisionlessly
at least until collision energies exceed the Planck scale.
As long as the freely-falling particles
remain focused along the null directions,
the energy-momenta will satisfy the hierarchy of conditions
\begin{equation}
\label{Thierarchy}
  T_{zz} \gg T_{za} \gg T_{aa}
  \ .
\end{equation}

The dominant energy-momenta are those $T_{zz}$ along the null directions.
Given these,
the Raychaudhuri equations~(\ref{Raychaudhuriexpansion})
and (\ref{Raychaudhuritwist}),
integrated along the outgoing and ingoing null directions,
yield the expansions and twists $\vartheta_z$ and $\varpi_z$
along the null directions.

The next largest energy-momenta are
the 4 radial-angular components $T_{za}$.
Given these, the Einstein
equations~(\ref{Einsteinza}),
integrated along the outgoing and ingoing null directions,
yield the expansions and twists $\vartheta_a$ and $\varpi_a$
along the angular directions.
These 4 Einstein equations essentially express
conservation of angular momentum.

The purely angular energy-momenta $T_{aa}$ are sub-sub-dominant.
The Einstein equations~(\ref{Einsteinaa}) for $T_{aa}$
depend on the angular expansions and twists $\vartheta_a$ and $\varpi_a$,
but since these
have already been determined by equations~(\ref{Einsteinza}) for $T_{za}$,
it should be that, if all is consistent,
the equations~(\ref{Einsteinaa}) for $T_{aa}$ should be satisfied identically.
If equations~(\ref{Einsteinaa}) are not satisfied
to adequate accuracy,
then it is a signal that the assumption of vanishing shear is breaking down.

The final two energy-momenta are $T_{vu}$ and $T_{+-}$.
As noted in \S\ref{equations},
if the energy-momenta $T_{kl}$ are arranged to satisfy
covariant energy-momentum conservation, as they should,
and if all is consistent,
then the Einstein equations for $T_{vu}$ and $T_{+-}$
should be satisfied automatically.

\section{Model line-element}
\label{modellineelement}

In this section I propose a particular line-element~(\ref{lineelement})
that I argue should,
during the early phases of inflation,
provide an adequate description
of the spacetime of a rotating black hole that accretes slowly,
and undergoes inflation at (just above) its inner horizon.

A physical motivation for the line-element~(\ref{lineelement})
is that the initial effect of inflationary energy-momentum,
which is focused along the principal null directions,
is to generate curvature (tidal accelerations) aligned with the principal frame.
The amount of curvature at any point varies
depending on the amplitude of the inflationary energy-momentum there,
which in turn depends on the accretion flow into the two principal directions
that pass through the point.
The incipient tidal deformation is encoded in the radial and angular
conformal factors $\rhox$ and $\rhoy$ in the line-element~(\ref{lineelement}).

The line-element~(\ref{lineelement}) is overly simple,
\S\ref{limitations},
but permits explicit expressions for the components $G_{vu}$ and $G_{+-}$
of the Einstein tensor,
equations~(\ref{Einsteinscalar}),
that clarify how inflation affects the evolution
of the conformal factors, \S\ref{evolution},
and provide the basis for some of the arguments in the next section,
\S\ref{physicalpicture}.

I refer to
the tetrad frame defined by the line-element~(\ref{lineelement})
as ``principal'' even though it is technically not so
(the complex self-dual Weyl tensor is not diagonal in this frame).
The frame is almost principal
in the sense that its null directions are the geodesic continuation
of the Kerr principal null directions,
and the spin~$\pm 2$ (but not spin~$\pm 1$)
components of the Weyl tensor vanish
(so the Petrov type is general, not Type D).

\subsection{Line-element}

The proposed line-element is essentially the Kerr line-element
modulated by two conformal factors,
a radial conformal factor $\rhox$ and an angular conformal factor $\rhoy$,
each of which could in principle be an arbitrary function of all 4 coordinates
$x$, $t$, $y$, $\phi$:
\begin{align}
\label{lineelement}
  \dd s^2
  &=
  \rhox^2
  \left[
  {\dd x^2 \over \Delta_x}
  -
  {\Delta_x \over ( 1 - \omega_x \omega_y )^2}
  \left( \dd t - \omega_y \dd \phi \right)^2
  \right]
\nonumber
\\
  &
  \ 
  +
  \rhoy^2
  \left[
  {\dd y^2 \over \Delta_y}
  +
  {\Delta_y \over ( 1 - \omega_x \omega_y )^2}
  \left( \dd \phi - \omega_x \dd t \right)^2
  \right]
  \ .
\end{align}
The expressions in square brackets are those of the Kerr line-element.
The coordinates $t$ and $\phi$
are time and azimuthal coordinates,
while the coordinates $x$ and $y$ are radial and angular coordinates
related to the usual Boyer-Linquist spheroidal coordinates $r$ and $\theta$ by
\begin{equation}
\label{rtheta}
  r
  \equiv
  a \cot ( a x )
  \ , \quad
  \cos\theta
  \equiv
  - y
  \ .
\end{equation}
The quantities
$\omega_x$ and $\omega_y$ are functions respectively
only of the radial and angular coordinates,
\begin{equation}
\label{omegaLKN}
  \omega_x
  =
  {a \over r^2 + a^2}
  \ , \quad
  \omega_y
  =
  a \sin^2\!\theta
  \ ,
\end{equation}
and likewise the radial and angular horizon functions
$\Delta_x$ and $\Delta_y$ are functions respectively
only of the radial and angular coordinates,
\begin{equation}
\label{DeltaLKN}
  \Delta_x
  =
  {1 \over r^2 + a^2}
  \left(
  1 - {2 \Mbh r \over r^2 + a^2}
  \right)
  \ , \quad
  \Delta_y
  =
  \sin^2\!\theta
  \ ,
\end{equation}
where $\Mbh$ is the black hole's mass.

The situation of interest is near the inner horizon,
where the radial horizon function
is negative and tending to zero, $\Delta_x \rightarrow - 0$.
The radial coordinate $x$ is then timelike
(and increasing inward, the direction of increasing proper time),
while the time coordinate $t$ is spacelike.

Since the black hole is slowly accreting,
at any given time, call it $t = 0$,
the line-element from just above the inner horizon upward
should closely approximate the Kerr line-element,
in which case the conformal factors $\rhox$ and $\rhoy$
are given by their Kerr value
\begin{equation}
\label{rhoLKN}
  \rhox
  =
  \rhoy
  =
  \rhosep
  =
  \sqrt{r^2 + a^2 \cos^2\!\theta}
  \ .
\end{equation}
Because the conformal factors $\rhox$ and $\rhoy$
are allowed to be functions of time $t$,
the black hole can grow.
From just above the inner horizon inward,
inflation accelerates the conformal factors away from the
Kerr value~(\ref{rhoLKN}).

The line-element~(\ref{lineelement})
generalizes the conformally separable line-element considered by
\cite{Hamilton:2010a,Hamilton:2010b,Hamilton:2010c}.
Specifically, the transformations (there $\rightarrow$ here)
\begin{equation}
  \rho \rightarrow \rhoy
  \ , \quad
  \dd x
  \rightarrow
  {\rho_x^2 \over \rho_y^2}
  \dd x
  \ , \quad
  \Delta_x
  \rightarrow
  {\rho_x^2 \over \rho_y^2}
  \Delta_x
  \ ,
\end{equation}
bring the conformally separable line-element of
\cite{Hamilton:2010a,Hamilton:2010b,Hamilton:2010c}
to the form~(\ref{lineelement}).
The advantage of this transformation is that,
if the conformal factors $\rhox$ and $\rhoy$ are allowed to be arbitrary,
then approximate solutions with arbitrarily time- and space-dependent accretion
flow can be admitted.
The solutions are no longer conformally separable,
so null geodesics are no longer exactly solvable,
but they prove do be solvable to adequate accuracy,
\S\ref{geodesics}.

The line-element~(\ref{lineelement})
defines not only
a metric
$g_{\kappa\lambda}$,
which is an inner product of coordinate tangent vectors
$\bg_\kappa$,
but also,
through
\begin{equation}
  \bg_\kappa \cdot \bg_\lambda
  =
  g_{\kappa\lambda}
  =
  \eta_{kl} e^k{}_\kappa e^l{}_\lambda
  =
  e^k{}_\kappa \bgamma_k \cdot e^l{}_\lambda \bgamma_l
  \ ,
\end{equation}
an inverse vierbein
$e^k{}_\kappa$,
and a corresponding orthonormal tetrad
$\{ \bgamma_x , \bgamma_t , \bgamma_y , \bgamma_\phi \}$,
whose inner products form the Minkowski metric,
$\bgamma_k \cdot \bgamma_l = \eta_{kl}$.
The Newman-Penrose tetrad
$\{ \bgamma_v , \bgamma_u , \bgamma_+ , \bgamma_- \}$
corresponding to the orthonormal tetrad is
\begin{equation}
  \bgamma_{\overset{\scriptstyle v}{\scriptstyle u}}
  \equiv
  {\textstyle \frac{1}{\sqrt{2}}}
  ( \bgamma_x \pm \bgamma_t )
  \ , \quad
  \bgamma_\pm
  \equiv
  {\textstyle \frac{1}{\sqrt{2}}}
  ( \bgamma_y \pm \im \bgamma_\phi )
  \ .
\end{equation}

\subsection{Geodesics}
\label{geodesics}

The low-density, hyper-relativistically counter-streaming streams
in the inflationary zone of a typically slowly accreting astronomical black hole
should behave collisionlessly,
at least until collision energies reach the Planck scale,
\S\ref{bhpa}.
Determining the collisionless energy-momentum requires solution of
the number density $N$ and tetrad-frame momentum $p_k$ along
freely-falling trajectories.
The energy-momentum tensor $T_{kl}$
will then be a sum over collisionless streams,
\begin{equation}
\label{Tkl}
  T_{kl}
  =
  \sum_{\rm streams}
  n_k p_l
  \ ,
\end{equation}
where $n^k$ is the number current
\begin{equation}
\label{nk}
  n^k
  =
  N
  p^k
  \ .
\end{equation}

Conformally separability is the proposition that
the equations of motion of massless particles are Hamilton-Jacobi separable.
The condition is weaker than strict separability
(that is,
Hamilton-Jacobi separability for massive as well as massless geodesics).
Whereas strict separability leads to electrovac and cognate spacetimes
\cite{Carter:1968c},
conformal separability, by allowing an arbitrary overall conformal factor,
also admits inflationary solutions
\cite{Hamilton:2010a,Hamilton:2010b,Hamilton:2010c}.

The line-element~(\ref{lineelement}) is not conformally separable,
because it has two arbitrary conformal factors instead of just one.
However,
under the conditions peculiar to inflation,
namely that motions are hyper-relativistic,
and the radial horizon function $\Delta_x$ is almost zero,
the spacetime is close enough to being conformally separable
that geodesics, massless or massive, are given adequately by
a Hamilton-Jacobi approximation.
Conformal separability and the accuracy of Hamilton-Jacobi separation
are discussed at length in \cite{Hamilton:2010b}.
Here I sketch only the main points
relevant to the present line-element~(\ref{lineelement}).

Write the covariant components $p_k$ of the tetrad-frame momentum
of a freely-falling particle of mass $m$
in terms of a set of Hamilton-Jacobi parameters $P_k$,
\begin{align}
  &\{ p_x , p_t , p_y , p_\phi \}
  =
\nonumber
\\
\label{pk}
  &
  \left\{
  {P_x \over \rhox \sqrt{- \Deltax}}
  ,
  {P_t \over \rhox \sqrt{- \Deltax}}
  ,
  {P_y \over \rhoy \sqrt{\Deltay}}
  ,
  {P_\phi \over \rhoy \sqrt{\Deltay}}
  \right\}
  \ ,
\end{align}
given by
\begin{subequations}
\label{Pk}
\begin{align}
  P_t
  &=
  \pi_t + \pi_\phi \omega_x
  \ ,
\\
  P_\phi
  &=
  \pi_\phi + \pi_t \omega_y
  \ ,
\\
\label{Px}
  P_x
  &=
  \sqrt{
  P_t^2 -
  (\rhox^2 / \rhoy^2)
  \left[ \KCarter + m^2 ( \rhoy^2 - a^2 \cos^2\!\theta ) \right]
  \Delta_x
  }
  \ ,
\\
  P_y
  &=
  \sqrt{
  - \, P_\phi^2 +
  \left( \KCarter - m^2 a^2 \cos^2\!\theta \right)
  \Delta_y
  }
  \ .
\end{align}
\end{subequations}
The constants $\pi_t$, $\pi_\phi$, and $\KCarter$
here are the conserved energy, azimuthal angular momentum,
and Carter constant of the particle in the parent Kerr spacetime
above the inner horizon.
The Hamilton-Jacobi parameters $P_k$ differ from the Kerr parameters
only in that $P_x$ involves the conformal factors $\rhox$ and $\rhoy$.
The Hamilton-Jacobi parameter $P_x$ itself is chosen
so that mass conservation is respected,
$p^k p_k = - m^2$.
That equations~(\ref{pk}) and (\ref{Pk}) provide an adequate approximation
is plausible from the fact that the horizon function $\Delta_x$
is tiny during inflation, so the factor proportional to $\Delta_x$
under the square root in the expression~(\ref{Px}) for $P_x$
is small, so the behavior of the conformal factors $\rhox$ and $\rhoy$
is irrelevant.
Conversely,
the Hamilton-Jacobi approximation~(\ref{pk}), (\ref{Pk})
can be expected to break down
when $(\rhox^2 / \rhoy^2) \Delta_x$ ceases to be small.
This occurs
when the angular conformal factor $\rhoy$ has collapsed to the point that
rotation becomes significant,
that is, angular motions $p_y$, $p_\phi$ become comparable
to radial motions $p_t$.

In the Hamilton-Jacobi approximation,
the number density $N$ along any collisionless stream satisfies
\cite{Hamilton:2010b}
\begin{equation}
\label{NHJ}
  N
  \propto
  {1 - \omega_x \omega_y \over \rhoy^2 P_x P_y}
  \ .
\end{equation}

Appendix~\ref{accuracyHJ} confirms that
the Hamilton-Jacobi approximations~(\ref{pk}), (\ref{Pk}), and (\ref{NHJ})
are adequately accurate,
as long as $(\rhox^2 / \rhoy^2) \Delta_x$ remains small.

\subsection{Expansions, twists, Einstein tensor}

The line-element~(\ref{lineelement})
satisfies the conditions~(\ref{nosheara}), (\ref{noshearz})
of vanishing shear along all 4 tetrad directions.
Consequently all the equations derived in \S\ref{completeequations}
hold for this line-element.
The expansions and twists
in the 4 tetrad directions are
\begin{subequations}
\begin{align}
\label{varthetaz}
  \vartheta_z
  &=
  \partial_z \ln \rhoy
  \ ,
\\
\label{varthetaa}
  \vartheta_a
  &=
  \partial_a \ln \rhox
  \ ,
\\
\label{varpiz}
  \varpi_v
  =
  \varpi_u
  &=
  -
  {\rhox \sqrt{- \Delta_x}
  \over
  2 \sqrt{2} \rhoy^2 ( 1 - \omega_x \omega_y )}
  {\dd \omega_y \over \dd y}
  \ ,
\\
\label{varpia}
  \varpi_+
  =
  \varpi_-
  &=
  {\rhoy \sqrt{\Delta_y}
  \over
  2 \sqrt{2} \rhox^2 ( 1 - \omega_x \omega_y )}
  {\dd \omega_x \over \dd x}
  \ ,
\end{align}
\end{subequations}
where $\partial_k$ denote tetrad-frame directed derivatives
(not coordinate-frame derivatives).

Explicit expressions for
the components of the Einstein tensor
not already given in \S\ref{equations},
namely $G_{vu}$ and $G_{+-}$,
can be expressed in terms of $\nu$ and $\mu$ defined by
\begin{equation}
\label{numu}
  \nu
  \equiv
  \ln \left(
  {1 - \omega_x \omega_y
  \over
  \rhox \sqrt{- \Delta_x}}
  \right)
  \ , \quad
  \mu
  \equiv
  \ln \left(
  {1 - \omega_x \omega_y
  \over
  \rhoy \sqrt{\Delta_y}}
  \right)
  \ .
\end{equation}
The quantity $\nu$ is physically the boost angle
(logarithm of the blueshift, or Lorentz $\gamma$-factor)
of affinely-parameterized principal null geodesics
relative to the tetrad frame.
Affinely-parameterized outgoing and ingoing principal null geodesics
are blueshifted by a factor $\ee^{2 \nu}$ relative to each other.
Similarly $\mu$
is the spatial rotation angle
of frames parallel-transported along the angular directions,
relative to the tetrad frame.
The components $G_{vu}$ and $G_{+-}$ of the Einstein tensor are
\begin{subequations}
\label{Einsteinscalar}
\begin{align}
  G_{vu}
  &=
  \DD^a \partial_a \mu
  -
  \DD^z \vartheta_z
  -
  \DD^a \vartheta_a
\nonumber
\\
\label{Guv}
  &\quad
  - \,
  2
  \vartheta^z \vartheta_z
  +
  2
  \varpi^z \varpi_z
  -
  \vartheta^a \vartheta_a
  -
  \varpi^a \varpi_a
  \ ,
\\
\label{Gpm}
  G_{+-}
  &=
  - \,
  \DD^z \partial_z \nu
  +
  \DD^z \vartheta_z
  +
  \DD^a \vartheta_a
\nonumber
\\
  &\quad
  + \,
  \vartheta^z \vartheta_z
  +
  \varpi^z \varpi_z
  +
  2
  \vartheta^a \vartheta_a
  -
  2
  \varpi^a \varpi_a
  \ .
\end{align}
\end{subequations}
In equations~(\ref{Einsteinscalar}),
as previously in equations~(\ref{Weylscalar}),
radial and angular vectors
are to be interpreted as
lying in their respective 2-dimensional tangent spaces.
Expressions~(\ref{Einsteinscalar})
for the Einstein components $G_{vu}$ and $G_{+-}$
generalize equations~(10) of \cite{Hamilton:2010a}.

\subsection{Evolution equations for the conformal factors}
\label{evolution}

During inflation,
infalling streams become highly focused along the
outgoing and ingoing principal null directions,
and they generate a large streaming energy-momentum
along those directions.
The Raychaudhuri equations~(\ref{RaychaudhuriSachs})
then imply that the expansions (but not twists)
along the null directions grow large.
For the line-element~(\ref{lineelement}),
the radial expansions are radial gradients of the angular conformal factor,
equation~(\ref{varthetaz}).
Thus during inflation radial gradients grow large,
and these dominate the expressions~(\ref{Einsteinscalar})
for $G_{vu}$ and $G_{+-}$.
The Einstein equations for the energy-momenta
$T_{vu}$ and $T_{+-}$
then reduce to
\begin{subequations}
\label{Einsteinvupm}
\begin{align}
\label{Einsteinvu}
  - \, \DD^z \vartheta_z - 2 \vartheta^z \vartheta_z
  &\approx
  8\pi T_{vu}
  \ ,
\\
\label{Einsteinpm}
  - \, \DD^z \partial_z \nu
  + \DD^z \vartheta_z
  + \vartheta^z \vartheta_z
  &\approx
  8 \pi T_{+-}
  \ .
\end{align}
\end{subequations}
The trace
$\vartheta^z \vartheta_z$
of the radial expansion is minus twice BIP's mass parameter.

As long as the outgoing and ingoing flows counter-stream collisionlessly,
the energy-momenta
$T_{vu}$ and $T_{+-}$
should be small,
so that the right hand sides of equations~(\ref{Einsteinvupm})
can be set to zero.
Equations~(\ref{Einsteinvupm})
with vanishing right hand sides
effectively generalize equations~(13a) and (13b) of \cite{Hamilton:2010a},
which formed the principal evolutionary equations of the latter paper.
Note that the latter paper,
elaborated by \cite{Hamilton:2010b},
was far more careful than the present paper.
By restricting to conformally separable solutions,
\cite{Hamilton:2010a}
was able to solve the Einstein equations by separation of variables
in a fashion that encompassed evolution
from electrovac through inflation to collapse down to a tiny scale.

In the inflationary (radial-gradient-dominated) regime,
the Raychaudhuri equation~(\ref{Sachsv}) for $T_{vv}$
and Einstein equation~(\ref{Einsteinvu}) for $T_{vu}$
can be written
\begin{subequations}
\label{Dvdrhoy}
\begin{align}
\label{Dvdvrhoy}
  \DD_v \partial_v \rhoy
  &\approx
  - \,
  4 \pi \rhoy T_{vv}
  \ ,
\\
\label{Dvdurhoy}
  \DD_v \partial_u \rhoy
  &\approx
  - \,
  {\Mass \over 2 \rhoy}
  +
  4 \pi r T_{vu}
  \ ,
\end{align}
\end{subequations}
where $\Mass$ is the mass function
\begin{equation}
\label{Mass}
  \Mass
  \equiv
  -
  \rhoy^2
  \vartheta^z \vartheta_z
  \ .
\end{equation}
In the case of spherical symmetry,
the corresponding exact Einstein equations are
\begin{subequations}
\label{Einsteinspher}
\begin{align}
\label{Einsteinvvspher}
  \DD_v \partial_v r
  &=
  - \,
  4 \pi r T_{vv}
  \ ,
\\
\label{Einsteinvuspher}
  \DD_v \partial_u r
  &=
  - \,
  {M \over r^2}
  +
  4 \pi r T_{vu}
  \ ,
\end{align}
\end{subequations}
where $r \equiv \rhoy$ is the circumferential radius,
and $M$ is the interior mass,
\begin{equation}
\label{Mspher}
  M
  \equiv
  {r^3 \over 2} \left(
  D^a \partial_a \mu
  - \vartheta^z \vartheta_z
  \right)
  \ .
\end{equation}
The spherically symmetric Einstein equation~(\ref{Einsteinvuspher})
differs from the radial-gradient-dominated Einstein equation~(\ref{Dvdurhoy})
by the inclusion of the sub-dominant term
$D^a \partial_a \mu = 1/r^2$
in the interior mass $M$, equation~(\ref{Mspher}).
Equations~(\ref{Einsteinspher})
are equivalent to the spherically symmetric
equations~(8) of
\cite{Hamilton:2008zz},
which constituted the principal evolutionary equations of the latter reference.

Thus
in the inflationary (radial-gradient-dominated) regime,
the equations governing the evolution of the radial expansions,
either~(\ref{Einsteinvupm}) or (\ref{Dvdrhoy}),
coincide
(modulo sub-dominant terms)
with the spherically symmetric equations.

With the expression~(\ref{varthetaz}) for the expansion $\vartheta_z$
and (\ref{numu}) for the boost $\nu$,
the Einstein equations~(\ref{Einsteinvupm}) combine to yield
\begin{equation}
\label{Drhox}
  \DD^z \partial_z \ln \left( {\rhox \rhoy^{1/2} \over \rhosep^{3/2}} \right)
  \approx
  4\pi \left( T_{vu} + 2 T_{+-} \right)
  \ .
\end{equation}
In deriving~(\ref{Drhox}),
use has been made of the fact that
$G_{vu}$ and $G_{+-}$, equations~(\ref{Einsteinscalar}),
vanish identically for Kerr,
and that the radial horizon function $\Deltax$ in $\nu$, equation~(\ref{numu}),
is the Kerr horizon function~(\ref{DeltaLKN})
(as long as $\Delta_x$ is a function only of $x$,
it can be set equal to the Kerr value
by a gauge transformation of the coordinate $x$
and the conformal factor $\rhox$).
If the flow is collisionless,
so that $T_{vu}$ and $T_{+-}$ remain small,
then,
given the boundary conditions that the conformal factors
are initially equal to the Kerr value, equation~(\ref{rhoLKN}),
equation~(\ref{Drhox}) integrates to
\begin{equation}
\label{rhox}
  {\rhox \rhoy^{1/2} \over \rhosep^{3/2}}
  \approx
  \mbox{constant}
  \ .
\end{equation}

\subsection{Limitations}
\label{limitations}

An obvious defect of
the line-element~(\ref{lineelement})
is that the Kerr parameters are fixed
(in comoving coordinates).
The arbitrary conformal factors allow the black hole to grow conformally,
so the black hole mass $\Mbh$ can grow
and the specific angular momentum along with it, $a \propto \Mbh$.
In reality however
the angular momentum of a black hole will evolve arbitrarily (but slowly)
in both magnitude and direction as the black hole accretes,
which conformal growth does not admit.

The slow variation of the Kerr parameters
of a slowly accreting black hole is potentially relevant to inflation because,
even though an outgoing or ingoing infaller experiences an extremely
short proper time during inflation,
they can see a considerable time go by on the highly blueshifted opposing stream
\cite{Hamilton:2008zz}.
An outgoing infaller sees ingoing particles accreted in the future,
while an ingoing infaller sees outgoing particles accreted in the past,
at times when the Kerr parameters of the black hole
could differ appreciably.

A second limitation of
the line-element~(\ref{lineelement})
is that it imposes that the radial expansion $\vartheta_z$
be pure gradient, equation~(\ref{varthetaz}).
The condition means that the expansion must be integrable:
the angular conformal factor $\rhoy$
can be obtained by integrating the expansions
along either of the outgoing and ingoing null directions,
and the same value of $\rhoy$ must be recovered
regardless of the integration path.
Consider the angular hypersurface inside the black hole
defined by the intersection of outgoing and ingoing null streams accreted
when the Kerr conformal factor of the black hole was respectively
$\rho_{\rm out}$ and $\rho_{\rm in}$.
As long as gravitational waves keep the slowly accreting black hole
close to hairless above its inner horizon,
the ratio $\rho_{\rm in} / \rho_{\rm out}$
is independent of angular position.
This condition is not a problem during early inflation
when the conformal factors are still near their initial Kerr value,
equation~(\ref{rhoLKN}),
but it becomes a serious constraint as inflation develops.
The integrability condition on the conformal factor $\rhoy$
translates into a condition that
the ingoing and outgoing fluxes $T_{vv}$ and $T_{uu}$
must balance in a fashion that is consistent over all angular positions.

The above defects both point to the fact that,
even if shear is in fact small,
gravitational waves cannot be neglected in a proper treatment of inflation
inside an arbitrarily accreting black hole.
It is gravitational waves that mediate the slow adjustment
of the parent Kerr black hole
from one angular momentum configuration to another.
And it is gravitational waves that keep the black hole close to hairless
above its inner horizon.

In practice these defects may after all be unimportant.
If Planck scale collisions intervene, as argued in \S\ref{bhpa},
then a stream sees only a few hundred black hole crossing times
elapse on the opposing stream before hitting the Planck wall.
This is short enough that the slow evolution of the parent Kerr geometry
should be unimportant.

\section{Physical picture}
\label{physicalpicture}

The physical picture that emerges from the arguments and equations
in \S\ref{gravitationalwaves}--\S\ref{modellineelement}
is consistent with that of BIP.

\subsection{Picture}
\label{picture}

The geodesics of the line-element~(\ref{lineelement}),
\S\ref{geodesics},
support the idea that
collisionless outgoing and ingoing streams remain focused along
the geodesic continuation of the principal null directions,
at least until the spacetime has collapsed to the point that
the intrinsic angular momentum of a stream
causes rotation to become important.

Inflationary energy-momentum focused along the principal null directions
produces a tidal gravitational acceleration
that tends to contract the angular conformal factor $\rhoy$
and elongate the radial conformal factor $\rhox$ in a 2 to 1 ratio,
equation~(\ref{rhox}),
\begin{equation}
  \rhox \propto \rhoy^{-1/2}
  \ .
\end{equation}

For a typically slowly accreting black hole,
the blueshift between outgoing and ingoing streams
approaching the inner horizon
grows so rapidly that the spacetime does not have time to react
to the enormously growing tidal acceleration.
During inflation, the conformal factors
$\rhox$ and $\rhoy$
are hugely accelerated (large second radial derivatives),
and start to change rapidly (large first radial derivatives),
but still the conformal factors themselves have hardly changed
from their initial Kerr values.
In effect,
inflation produces an explosion of acceleration
so rapid that it initially leaves the Kerr spacetime unchanged.
The fact that volume elements are initially little distorted
despite the enormous tidal acceleration
was first pointed out by \cite{Ori:1991}.
The enormous (in due course super-Planckian) tidal force
yet little distortion is commonly characterized as a
``weak null singularity.''


If a black hole continues to accrete,
as is always true in reality,
and if the issue of super-Planckian curvature is set aside,
then the end result of inflationary acceleration is collapse,
not a null singularity
\cite{Hamilton:2008zz}.
Whether the acceleration causes inflation or collapse
can be read off from the Einstein equations~(\ref{Dvdrhoy}).
These equations are essentially the same as those governing inflation
in spherical black holes,
equations~(8) of
\cite{Hamilton:2008zz}.
As argued in that paper,
collisionless counter-streaming tends to drive
exponentially growing acceleration,
while mass (the $\Mass$ term in equation~(\ref{Dvdurhoy}))
tends to cause collapse.
The reader is referred to
\cite{Hamilton:2008zz}
for a more detailed exposition.

\subsection{Is inflation similar to that in spherical symmetry?}

BIP advanced ``tentative'' arguments
that inflation in rotating black holes would be similar
to that in spherically symmetric black holes.
Is it true?

From the perspective of an infaller,
inflation takes place over an extremely short proper time.
Consequently infallers belonging to the same (outgoing or ingoing) stream
are causally connected over only an extremely short proper distance.
During inflation,
the blueshift of the opposing stream increases exponentially,
and the proper time decreases inversely with the blueshift.
An infaller sees approximately one black hole crossing time
elapse on the opposing stream for each $e$-fold of blueshift
\cite{Hamilton:2008zz}.
Thus the evolution of the spacetime
in any angular patch is determined only by the causal domain of that patch,
which becomes increasingly narrow as inflation develops.

The results of \S\ref{modellineelement}
support the idea that inflation in any angular patch
behaves as in spherical symmetry.
During early inflation,
inflationary energy-momentum at any point
produces an enormous tidal acceleration
that depends only on the energy-momentum passing through that point.
The tidal acceleration increases exponentially while
the underlying volume element remains scarcely distorted.
Once inflation gives way to collapse,
it is plausible that 
volume elements will simply react as they have been accelerated:
they will collapse in the angular direction
and stretch in the radial direction.
In this picture,
the behavior in any angular patch is entirely local to that patch.

The effect of departures from uniform angular behavior
can be assessed from the
Raychaudhuri-Einstein equations~(\ref{Einsteinza}),
which determine how the angular momentum of accreting matter
affects the angular momentum of the black hole.
Note that, as mentioned in \S\ref{limitations},
the line-element~(\ref{lineelement})
is inadequate to follow the evolution of the angular momentum of the black hole.
Applied to the line-element~(\ref{lineelement}),
the Raychaudhuri-Einstein equations~(\ref{Einsteinza})
merely force the accreting matter to have an angular momentum
consistent with that of the black hole.

Departures from uniform angular behavior arise from two causes:
first,
accretion with angular momentum per unit mass mismatched,
in amplitude or direction, from that of the black hole;
and second,
accretion at different rates at different angular positions.
The two causes provide two of the source terms,
namely
$T_{za}$
and
$\DD_a \Gamma_{azb}$,
in the Raychaudhuri-Einstein equations~(\ref{Einsteinza})
governing the evolution
$\DD_z \Gamma_{azy}$ of the angular expansions and twists.

Since angular energy-momenta $T_{za}$
are sub-dominant to radial energy-momenta $T_{zz}$,
order unity differences in the angular energy-momenta
should have little effect on the dominant radial evolution of the spacetime
(the radial tidal distortion).
That the effect should be small is supported by
the solutions of
\cite{Hamilton:2010a,Hamilton:2010b},
in which inflation was followed by collapse,
and rotation became important only after the geometry had collapsed
to an exponentially tiny scale.
Since the specific angular momentum of accreted matter
cannot be much larger than that of the black hole,
otherwise the matter would not be accreted,
it follows that the angular momentum
$T_{za}$
cannot be large enough to affect radial evolution substantially.

Accretion at different rates at different angular positions
could potentially have a larger effect.
Larger accretion rates cause more rapid collapse.
Order unity angular gradients in the incident accretion flow
translate into order unity differences across a causal angular patch.
As with the accreted angular momentum $T_{za}$,
order unity differences should not have much effect.
But it is easy to think of situations,
such as where accretion is confined to delta-functions of angular location,
where angular gradients are large.

Gravitational waves are likely to be generated most strongly
where accretion is most anisotropic.
Gravitational waves should effectively spread out the accretion flow,
by transforming some of the accretion energy
into a collisionless fluid of gravitons more uniformly distributed
over the inner horizon of the black hole.
Thus gravitational waves should set an upper limit on anisotropy.
It is not clear that the large angular gradients needed
to produce significant departures from quasi-spherical collapse
can be sustained in the presence of smoothing by gravitational waves.

While these arguments are not conclusive,
they tend to support BIP's proosal that
angular anisotropy plays a sub-dominant role
in inflation in rotating black holes.



\subsection{Ori}

In one of the few studies of inflation in rotating black holes, Ori
\cite{Ori:1992,Ori:2001pc}
considered the situation of a rotating black hole that collapses
and then remains isolated,
in which case a Price tail of gravitational radiation
provides the outgoing and ingoing streams that drive inflation.
To the extent that gravitational waves can be treated as
a collisionless fluid of gravitons,
the situation considered by Ori can be compared to the
results of the present paper.

Ori argued that perturbations of the geometry in the inflationary
zone at the inner horizon would have some simplifying features.
This is at least qualitatively consistent with two simplifying
features of inflation noted in the present paper,
namely that geodesics focus along the principal null directions
of the parent Kerr geometry regardless of their orbital parameters,
and that the effect of inflation is to generate large accelerations
that leave the underlying Kerr spacetime at least initially unchanged.

Ori's principal conclusion was that
the resulting null singularity in a rotating black hole
would differ from that in a spherical black hole
by being oscillatory.
The oscillations would be dominated by quadrupole radiation,
the lowest order of gravitational waves.
The natural interpretation of this conclusion
in the context of the present paper
is that the quadrupole oscillations are the tail end of the
Price tail of gravitational radiation witnessed by an infaller.
Although the gravitational waves have a quadrupole angular pattern
over the black hole,
an infaller will see these waves hugely blueshifted,
and concentrated into an intense, narrow beam
focused along the opposing principal null direction.

\section{Planck wall}
\label{bhpa}

At what point does the classical general relativistic
description of inflation fail because of quantum gravitational effects?
\cite{Hamilton:2008zz}
pointed out that in a typical astronomical black hole,
Planck-scale physics is first encountered in
collisions between outgoing and ingoing particles,
which reach Planck center-of-mass energies
well before the curvature hits the Planck scale.
The purpose of this section is to explore this idea
to see where it leads.
The units in this section are Planck units,
$c = G = \hbar = 1$.

It is widely expected that collision cross-sections
will become gravitational at super-Planckian energies
and therefore increase rapidly with energy,
increasing approximately as the horizon radius squared
of a black hole whose mass equals the center-of-mass energy
\cite{Banks:1999gd,Giddings:2001bu,Dimopoulos:2001hw,Giddings:2006vu}.
By contrast,
total cross-sections for non-gravitational processes
are dominated by small-angle scattering,
and increase more slowly.
Total cross-sections for electromagnetic inelastic scattering
(bremsstrahlung)
increase no faster than logarithmically with energy
\cite{Matthews1973157,Seltzer198595}.
Total cross-sections for short-range interactions,
such as the nuclear force,
are limited by unitarity to increasing no faster than
the logarithm squared of the center-of-mass energy, the Froissart bound
\cite{Froissart:1961ux,Martin:1963,Martin:1965,Block:2006hy,Martin:2009pt}.
In the absence of collisions,
inflation would easily accelerate outgoing and ingoing particles
far above Planck center-of-mass energies.
Thus if collision cross-sections do increase rapidly
above the Planck scale,
then inevitably inflation will reach a ``Planck wall''
where collisions become important,
and presumably lead to a burst of entropy production.

According to the arguments of
\cite{Hamilton:2008zz},
if collisions convert sufficient streaming energy
into center-of-mass energy,
then exponential inflationary growth will cease,
and the spacetime will collapse.




The blueshift of either of the outgoing or ingoing streams
relative to the center-of-mass frame
is $\ee^{\nu}$.
For brevity denote this blueshift by $u$,
\begin{equation}
 u \equiv
 \ee^{\nu}
 \ .
\end{equation}
During inflation, the blueshift $u$ increases exponentially.
By the estimate of
\cite{Hamilton:2010b},
in the absence of collisions,
and for an accretion rate of $\Mbhdot$,
inflation gives way to collapse
when the blueshift reaches the maximum value
\begin{equation}
  u_{\rm max}
  \approx
  \ee^{-1 / \Mbhdot}
  \ .
\end{equation}
For a typically slowly accreting black hole,
the accretion rate is small,
$\Mbhdot \ll 1$,
and the maximum blueshift $u_{\rm max}$ is exponentially huge.
For collision energies to reach the Planck scale before collapse,
the accretion rate must be smaller than about $0.01$,
\begin{equation}
  \Mbhdot
  \lesssim
  0.01
  \ .
\end{equation}
If the accretion rate is larger,
such as may happen when the black hole first collapses,
or during a black hole merger,
then the spacetime will collapse before the blueshift inflates
to the Planck scale,
and the situation envisaged in this section will not occur.

The proper time $\tau$ experienced
in the center-of-mass frame
as the blueshift increases by one $\ee$-fold is
the black hole crossing time $\Mbh$ divided by the blueshift $u$,
\begin{equation}
\label{tauinf}
  \tau \approx \Mbh / u
  \ .
\end{equation}
The proper time $\tau_{\rm coll}$
in the center-of-mass frame
for an outgoing or ingoing particle to
collide with a particle in the opposing stream is
\begin{equation}
\label{taucoll}
  \tau_{\rm coll}
  \approx
  {1 \over n u \sigma}
  \approx
  {m \Mbh^2 \over \Mbhdot u \sigma}
  \ ,
\end{equation}
where $n$ is the proper number density of particles in the opposing stream
(in the latter's own frame),
and $\sigma$ is the collision cross-section.
The proper number density $n$ equals the accretion rate
$\Mbhdot$,
multiplied by the density of the black hole, which is $\Mbh^{-2}$,
divided by the rest mass $m$ of the particle,
$n \approx \Mbhdot / ( \Mbh^2 m )$.
The collision time~(\ref{taucoll})
is less than an inflation time~(\ref{tauinf})
if the cross-section $\sigma$ exceeds the critical cross-section
$\sigma_c$ given by
\begin{equation}
\label{sigmac}
  \sigma_c
  \approx
  {m \Mbh \over \Mbhdot}
  \approx
  {10 \unit{fb} \over \Mbhdot}
  \left( {m \over 1 \unit{TeV}} \right)
  \left( {\Mbh \over 10^6 \unit{\Msun}} \right)
  \ ,
\end{equation}
where a femtobarn $1 \unit{fb} = 10^{-43} \unit{m}^2$
is about a weak interaction cross-section.

Nature will provide a broad range of accretion rates $\Mbhdot$.
In a ``typical'' situation where the accretion rate
is small but not too small,
the collision time will be shorter than an inflation time
for electromagnetic and strong interactions,
but longer than an inflation time for weak interactions.
In this situation,
collisions will keep baryons, electrons, and photons tightly coupled,
forcing them into a common outgoing or ingoing stream
before inflation ignites.
Magnetohydrodynamic processes will contribute to keeping
the baryonic plasma tightly coupled
\cite{Balbus:1998,Balbus:2003xh}.
On the other hand particles that interact only
by weak interactions or gravity, such as dark matter particles or gravitons,
can occupy the opposing stream,
and stream relativistically through the baryonic stream without collisions,
driving inflation.

If the accretion rate $\Mbhdot$ is large enough,
and if dark matter particles are weakly interacting,
then electroweak-scale collisions
could lead to significant entropy production.
Currently only upper limits to cross-sections between
dark matter particles and baryons are known
\cite{Desai:2004pq,Akerib:2004fq,Abbasi:2009uz}.

Since wavelengths decrease inversely with momentum,
unknown processes with thresholds at higher (sub-Planckian) energies
will probably have correspondingly smaller cross-sections,
and will probably not contribute significantly to collisions.
Given the slow increase of non-gravitational cross-sections with energy,
inflation will typically blueshift up to the Planck energy without
meeting a major collisional barrier.

Above a Planck energy, cross-sections increase gravitationally.
The Planck wall will be reached when the gravitational cross-section
exceeds the critical cross-section~(\ref{sigmac}).
Notwithstanding the lack of a robust theory of quantum gravity,
the common expectation is that super-Planckian collisions
will lead to the production of mini black holes
\cite{Banks:1999gd,Giddings:2001bu,Dimopoulos:2001hw,Giddings:2006vu,Dai:2007ki,Choptuik:2009ww,Gingrich:2009hj,Witek:2010xi,Witek:2010az}.
Hereafter I will refer to the product of a super-Planckian collision
as a proto mini black hole, or pmbh.
If the Planck energy is at the standard value $10^{16} \unit{TeV}$,
then the mass of a pmbh at the Planck wall is
$\mbh \approx \sqrt{\sigma_c}$, or
\begin{equation}
\label{mbh}
  \mbh
  \approx
  \sqrt{{m \Mbh \over \Mbhdot}}
  \approx
  {10^{14} \over \sqrt{\Mbhdot}}
  \left( {m \over 1 \unit{TeV}} \right)^{1/2}
  \left( {\Mbh \over 10^{6} \unit{\Msun}} \right)^{1/2}
  \ .
\end{equation}
Modulo a factor of $\Mbhdot^{-1/2}$,
the mass $\mbh$ of the pmbh is
the geometric mean
of the masses $m$ and $\Mbh$
of the colliding particle and the parent black hole.
The collision energy of
$\mbh \gtrsim 10^{14}$ Planck masses
is large compared to any collision process one could imagine
happening elsewhere in the Universe.

At the Planck wall,
the blueshift $u_{\rm wall}$ is
\begin{equation}
  u_{\rm wall}
  \approx
  {\mbh \over m}
  \approx
  {10^{30} \over \sqrt{\Mbhdot}}
  \left( {m \over 1 \unit{TeV}} \right)^{-1/2}
  \left( {\Mbh \over 10^{6} \unit{\Msun}} \right)^{1/2}
  ,
\end{equation}
and the proper time
$\tau_{\rm wall}$
for the blueshift to increase one $\ee$-fold is
\begin{align}
\label{tauwall}
  \tau_{\rm wall}
  &\approx
  {\Mbh \over u_{\rm wall}}
  \approx
  \mbh \Mbhdot
\nonumber
\\
  &\approx
  10^{14} \sqrt{\Mbhdot}
  \left( {m \over 1 \unit{TeV}} \right)^{1/2}
  \left( {\Mbh \over 10^{6} \unit{\Msun}} \right)^{1/2}
  \ ,
\end{align}
which is larger than a Planck time
if the accretion rate is not too tiny,
$\Mbhdot \gtrsim 10^{-28}$.
Regardless of whether inflation continues or gives way to collapse,
the time $\tau_{\rm wall}$
sets the characteristic proper time remaining
before the background spacetime comes to an end
(or at least, the spacetime reaches Planck density).
The proper time $\tau_{\rm wall}$,
equation~(\ref{tauwall}),
is less than of the order of
the crossing time $\mbh$ of the pmbh.
Thus the pmbh does not have time to express itself as a black hole,
which takes several crossing times.
The pmbh certainly does not have time to evaporate by Hawking radiation,
which takes a time $\mbh^3$.
Meanwhile, the mass density of the background spacetime is
\begin{equation}
  m n u_{\rm wall}^2
  \approx
  {\Mbhdot \over \Mbh m}
  \approx
  {1 \over \mbh^2}
  \ ,
\end{equation}
which is already the density of the putative pmbh.

Does a collision actually occur,
if the pmbh does not have time to express itself as a black hole?
Yes.
In the center-of-mass frame,
the colliding outgoing and ingoing particles are Lorentz-contracted to size
$1 / ( m u )$,
and it takes that proper time
for the particles to pass by each other,
\begin{equation}
  \tau_{\rm pass}
  \approx
  {1 \over \mbh}
  \ ,
\end{equation}
which is short compared to the time $\tau_{\rm wall}$ available.

The fact that super-Planckian collisions do occur
indicates that entropy production takes place,
but the fact that there is not enough time for the
collision products to express themselves as mini black holes
indicates that the situation is messy.
One possibility is that entropy production might lead to a
stringy Hagedorn phase
\cite{Nayeri:2005ck,Brandenberger:2011et}.

String theory predicts the existence of higher dimensions.
It has been speculated that if some of the extra dimensions are large,
then the fundamental Planck energy could be as low as $1 \unit{TeV}$
\cite{ArkaniHamed:1998nn},
and that super-Planckian collisions could create higher-dimensional
mini black holes with radii between the fundamental Planck scale
and the scale of the large extra dimensions
\cite{Banks:1999gd,Giddings:2001bu,Dimopoulos:2001hw,Giddings:2006vu}.
If this scenario is correct,
then the critical cross-section~(\ref{sigmac})
still sets the radius of pmbhs at the Planck wall,
but their mass $\mbh$ will be smaller
(though still necessarily super-Planckian).
Their dimensionality will be the number of dimensions
larger than the radius of the critical cross-section.
The possibility of extra dimensions does not change the conclusion
that inflation will meet a Planck wall
where collisions become important.


Before hitting the Planck wall,
the inflationary zone of the black hole
behaves like a particle accelerator,
accelerating twin beams of particles through each other,
and conducting numerous collision experiments at energies up to $\mbh$.
Even highly improbable collision events might occur.

\section{Conclusions}

In this paper I have presented a number of proposals
about how inflation should develop inside a rotating black hole
accreting slowly but in an arbitrary time- and space-dependent fashion.

As shown by
\cite{Hamilton:2010b},
particles freely-falling in the parent Kerr geometry of a rotating black hole
become highly focused
along the principal outgoing and ingoing null directions
as they approach the inner horizon,
regardless of their initial orbital parameters.
If the Kerr geometry remained unchanged,
then the streaming energy of the outgoing and ingoing streams would
diverge at the inner horizon.
In practice,
the energy-momentum focused along the principal null directions
produces a tidal acceleration that tends to collapse angular directions
and stretch radial directions.
For a slowly accreting black hole,
the proper timescale over which the streaming energy grows is so short
that the tidal acceleration grows exponentially huge
before volume elements actually begin to distort,
as first pointed out by \cite{Ori:1991}.
In due course the enormous acceleration leads to collapse
of the angular directions.
Because of the extremely short proper time,
angular patches are causally connected over extremely short proper distances,
so the acceleration and collapse of an angular patch
depends only on the streaming energy local to that patch.

The arguments of the present paper generally support
BIP's
\cite{Barrabes:1990}
proposal that inflation in rotating black holes
would evolve similarly to that in spherical black holes.
Yet it is surprisingly difficult to demonstrate this proposal conclusively.
In \S\ref{gravitationalwaves}
I argue that during inflation
gravitational waves should behave like a collisionless fluid of gravitons,
and consequently that the spacetime should be almost shear-free,
as it is in the parent Kerr geometry.
But even if the spacetime is shear-free,
gravitational waves still play a crucial role
both in keeping the parent
Kerr black hole almost hairless as it evolves slowly from one
angular momentum configuration to another,
and in smoothing the angular distribution of accretion energy
when the accretion flow is highly anisotropic.

It would be desirable to test the proposals of the present paper
with full 4-dimensional numerical simulations.
Such simulations will not be easy.
One challenge is that
inflation generates enormous spatial and temporal gradients.
A second challenge is the need to model outgoing and ingoing streams
that counter-stream relativistically.
A third challenge is how to treat gravitational waves
both in the familiar regime where they radiate away a black hole's hair,
and in the inflationary regime where they behave effectively like
a collisionless fluid.
A final difficulty in comparing theory to simulation is that
theoretical arguments are simplest when accretion rates are small,
but numerical simulations should be easiest in the opposite regime,
when accretion rates are high.
It may help to work in a double-null 2+2 formalism,
since the physics in any angular patch is that of two effectively null streams
streaming through each other.
Since angular patches become causally disconnected from each other
as inflation progresses,
it may help to confine the simulations to one causal patch.

As pointed out by
\cite{Hamilton:2008zz},
center-of-mass collision energies between outgoing and ingoing particles
will hit the Planck energy well before
the curvature reaches the Planck scale.
It is thought that collision cross-sections will become gravitational,
and therefore increase rapidly, at super-Planckian collision energies
\cite{Banks:1999gd,Giddings:2001bu,Dimopoulos:2001hw,Giddings:2006vu}.
If so, then inflation will reach a ``Planck wall''
where the collision time is less than the inflation time.
In a typical supermassive black hole,
the collision energy at the Planck wall is $\gtrsim 10^{14}$ Planck masses,
equation~(\ref{mbh}).
It is thought that super-Planckian collisions will lead to
the formation of mini black holes
\cite{Banks:1999gd,Giddings:2001bu,Dimopoulos:2001hw,Giddings:2006vu},
but that does not happen here because the timescale for the spacetime
to collapse is less than the size of a nascent mini black hole.
Collisions do actually occur,
since the collapse timescale is easily long enough to allow
two colliding particles to pass by each other.
Presumably collisions lead to entropy production,
but the situation appears complicated.
Qualitatively,
entropy production should bring inflation to an end and precipitate collapse
\cite{Hamilton:2008zz}.
It is possible that
collisions might lead to a stringy Hagedorn phase
\cite{Brandenberger:2011et}.

Prior to hitting the Planck wall,
inflation acts like a particle accelerator of extraordinary power.
It appears inescapable that
Nature is conducting vast numbers of collision experiments
over a broad range of peri- and super-Planckian energies
in large numbers of black holes throughout our Universe.
Does Nature do anything interesting with this extravagance
-- such as create baby universes --
or is it merely a final hurrah en route to nothingness?

\begin{acknowledgements}
I thank Gavin Polhemus for many helpful conversations.
This work was supported by NSF award
AST-0708607.
\end{acknowledgements}

\section*{References}

\bibliographystyle{unsrt}
\bibliography{bh}

\begin{thebibliography}{10}

\bibitem{Senovilla:1997}
Jos{\'e} M.~M. Senovilla.
\newblock Singularity theorems and their consequences.
\newblock {\em Gen.\ Rel.\ Grav.}, 29:701--848, 1997.

\bibitem{Poisson:1989zz}
E.~Poisson and W.~Israel.
\newblock Inner-horizon instability and mass inflation in black holes.
\newblock {\em Phys. Rev. Lett.}, 63:1663--1666, 1989.

\bibitem{Poisson:1990eh}
E.~Poisson and W.~Israel.
\newblock Internal structure of black holes.
\newblock {\em Phys. Rev.}, D41:1796--1809, 1990.

\bibitem{Penrose:1968}
Roger Penrose.
\newblock Structure of space-time.
\newblock In C{\'e}cile de~Witt-Morette and John~A. Wheeler, editors, {\em
  Battelle {R}encontres: 1967 lectures in mathematics and physics}, pages
  121--235. W. A. Benjamin, New York, 1968.

\bibitem{Hamilton:2008zz}
Andrew J.~S. Hamilton and Pedro~P. Avelino.
\newblock The physics of the relativistic counter-streaming instability that
  drives mass inflation inside black holes.
\newblock {\em Phys. Rept.}, 495:1--32, 2010.

\bibitem{Chan:1994rs}
J.~S.~F. Chan, K.~C.~K. Chan, and Robert~B. Mann.
\newblock Interior structure of a charged spinning black hole in
  (2+1)-dimensions.
\newblock {\em Phys. Rev.}, D54:1535--1539, 1996.

\bibitem{Barrabes:1990}
C.~Barrab{\`e}s, W.~Israel, and E.~Poisson.
\newblock Collision of light-like shells and mass inflation in rotating black
  holes.
\newblock {\em Class. Quant. Grav.}, 7(12):L273--L278, 1990.

\bibitem{Ori:1992}
Amos Ori.
\newblock Structure of the singularity inside a realistic rotating black hole.
\newblock {\em Phys. Rev. Lett.}, 68:2117--2121, 1992.

\bibitem{Ori:2001pc}
Amos Ori.
\newblock Oscillatory null singularity inside realistic spinning black holes.
\newblock {\em Phys. Rev. Lett.}, 83:5423--5426, 1999.

\bibitem{Hamilton:2010a}
Andrew J.~S. Hamilton and Gavin Polhemus.
\newblock The interior structure of rotating black holes 1. {S}ummary.
\newblock arXiv:1010.1269, 2010.

\bibitem{Hamilton:2010b}
Andrew J.~S. Hamilton.
\newblock The interior structure of rotating black holes 2. {U}ncharged black
  holes.
\newblock arXiv:1010.1271, 2010.

\bibitem{Hamilton:2010c}
Andrew J.~S. Hamilton.
\newblock The interior structure of rotating black holes 3. {C}harged black
  holes.
\newblock arXiv:1010.1272, 2010.

\bibitem{Ori:1991}
Amos Ori.
\newblock Inner structure of a charged black hole: an exact mass-inflation
  solution.
\newblock {\em Phys. Rev. Lett.}, 67:789--792, 1991.

\bibitem{Penrose:1965}
Roger Penrose.
\newblock Gravitational collapse and spacetime singularities.
\newblock {\em Phys. Rev. Lett.}, 14:57--59, 1965.

\bibitem{Belinski:1970}
Vladimir~A. Belinsky, Isaak~M. Khalatnikov, and Evgeny~M. Lifshitz.
\newblock Oscillatory approach to a singular point in the relativistic
  cosmology.
\newblock {\em Advances in Physics}, 19:525--573, 1970.

\bibitem{Belinski:1971}
Vladimir~A. Belinsky and Isaak~M. Khalatnikov.
\newblock General solution of the gravitational equations with a physical
  oscillatory singularity.
\newblock {\em Sov. Phys. JETP}, 32:169--172, 1971.

\bibitem{Belinski:1982}
Vladimir~A. Belinsky, Isaak~M. Khalatnikov, and Evgeny~M. Lifshitz.
\newblock A general solution of the {E}instein equations with a time
  singularity.
\newblock {\em Advances in Physics}, 31:639--667, 1982.

\bibitem{Berger:2002st}
Beverly~K. Berger.
\newblock Numerical approaches to spacetime singularities.
\newblock {\em Liv. Rev. Rel.}, 5:1, 2002.

\bibitem{Ashtekar:2011ck}
Abhay Ashtekar, Adam Henderson, and David Sloan.
\newblock A {H}amiltonian formulation of the {BKL} conjecture.
\newblock {\em Phys. Rev.}, D83:084024, 2011.

\bibitem{Price:1972}
Richard~H. Price.
\newblock Nonspherical perturbations of relativistic gravitational collapse.
  {I}. {S}calar and gravitational perturbations.
\newblock {\em Phys.\ Rev.}, 5:2419--2438, 1972.

\bibitem{Carter:1968c}
Brandon Carter.
\newblock Hamilton-{J}acobi and {S}chr{\"o}dinger separable solutions of
  {E}instein's equations.
\newblock {\em Commun.\ Math.\ Phys.}, 10:280--310, 1968.

\bibitem{Banks:1999gd}
Tom Banks and Willy Fischler.
\newblock A model for high energy scattering in quantum gravity.
\newblock hep-th/9906038, 1999.

\bibitem{Giddings:2001bu}
Steven~B. Giddings and Scott~D. Thomas.
\newblock High energy colliders as black hole factories: The end of short
  distance physics.
\newblock {\em Phys. Rev.}, D65:056010, 2002.

\bibitem{Dimopoulos:2001hw}
Savas Dimopoulos and Greg~L. Landsberg.
\newblock Black holes at the {LHC}.
\newblock {\em Phys. Rev. Lett.}, 87:161602, 2001.

\bibitem{Giddings:2006vu}
Steven~B. Giddings.
\newblock Locality in quantum gravity and string theory.
\newblock {\em Phys. Rev.}, D74:106006, 2006.

\bibitem{Matthews1973157}
J.~L. Matthews and R.~O. Owens.
\newblock Accurate formulae for the calculation of high energy electron
  bremsstrahlung spectra.
\newblock {\em Nuclear Instruments and Methods}, 111(1):157--168, 1973.

\bibitem{Seltzer198595}
Stephen~M. Seltzer and Martin~J. Berger.
\newblock Bremsstrahlung spectra from electron interactions with screened
  atomic nuclei and orbital electrons.
\newblock {\em Nuclear Instruments and Methods in Physics Research Section B:
  Beam Interactions with Materials and Atoms}, 12(1):95--134, 1985.

\bibitem{Froissart:1961ux}
Marcel Froissart.
\newblock Asymptotic behavior and subtractions in the {M}andelstam
  representation.
\newblock {\em Phys. Rev.}, 123:1053--1057, 1961.

\bibitem{Martin:1963}
A.~Martin.
\newblock Unitarity and high-energy behavior of scattering amplitudes.
\newblock {\em Phys. Rev.}, 129(3):1432--1436, Feb 1963.

\bibitem{Martin:1965}
A.~Martin.
\newblock Extension of the axiomatic analyticity domain of scattering
  amplitudes by unitarity-{I}.
\newblock {\em Nuovo Cimento}, 42(4):930--953, Apr 1965.

\bibitem{Block:2006hy}
Martin~M. Block.
\newblock Hadronic forward scattering: Predictions for the {L}arge {H}adron
  {C}ollider and cosmic rays.
\newblock {\em Phys. Rept.}, 436:71--215, 2006.

\bibitem{Martin:2009pt}
Andre Martin.
\newblock The {F}roissart bound for inelastic cross-sections.
\newblock {\em Phys.Rev.}, D80:065013, 2009.

\bibitem{Balbus:1998}
Steven~A Balbus and John~F. Hawley.
\newblock Instability, turbulence, and enhanced transport in accretion disks.
\newblock {\em Rev.\ Mod.\ Phys.}, 70:1--53, 1998.

\bibitem{Balbus:2003xh}
Steven~A. Balbus.
\newblock Enhanced angular momentum transport in accretion disks.
\newblock {\em Ann. Rev. Astron. Astrophys.}, 41:555--597, 2003.

\bibitem{Reno:2001hv}
Mary~Hall Reno, Ina Sarcevic, George~F. Sterman, Marco Stratmann, and Werner
  Vogelsang.
\newblock Ultrahigh energy neutrinos, small x and unitarity.
\newblock 2001.
\newblock hep-ph/0110235.

\bibitem{Basu:2002uu}
Rahul Basu, Debajyoti Choudhury, and Swapan Majhi.
\newblock {NLO} corrections to ultra-high energy neutrino nucleon scattering,
  saturation and small x.
\newblock {\em JHEP}, 10:012, 2002.

\bibitem{Sarcevic:2005cz}
I.~Sarcevic.
\newblock High energy neutrino interactions.
\newblock 2005.
\newblock hep-ph/0508002.

\bibitem{Henley:2005ms}
Ernest~M. Henley and Jamal Jalilian-Marian.
\newblock Ultra-high energy neutrino-nucleon scattering and parton
  distributions at small x.
\newblock {\em Phys.Rev.}, D73:094004, 2006.

\bibitem{Illarionov:2011wc}
Alexey~Yu. Illarionov, Bernd~A. Kniehl, and Anatoly~V. Kotikov.
\newblock Ultrahigh-energy neutrino-nucleon deep-inelastic scattering and the
  {F}roissart bound.
\newblock {\em Phys. Rev. Lett.}, 106:231802, 2011.

\bibitem{Choudhury:2011zj}
D~K Choudhury and Pijush~Kanti Dhar.
\newblock {DGLAP} evolutions and cross-sections of neutrino-nucleon interaction
  at ultra high energy.
\newblock 2011.
\newblock arXiv:1103.3788.

\bibitem{Desai:2004pq}
S.~Desai et~al.
\newblock Search for dark matter {WIMP}s using upward through-going muons in
  {S}uper-{K}amiokande.
\newblock {\em Phys. Rev.}, D70:083523, 2004.

\bibitem{Akerib:2004fq}
D.~S. Akerib et~al.
\newblock First results from the cryogenic dark matter search in the {S}oudan
  underground lab.
\newblock {\em Phys. Rev. Lett.}, 93:211301, 2004.

\bibitem{Abbasi:2009uz}
R.~Abbasi et~al.
\newblock Limits on a muon flux from neutralino annihilations in the {S}un with
  the {I}ce{C}ube 22-string detector.
\newblock {\em Phys. Rev. Lett.}, 102:201302, 2009.

\bibitem{Dai:2007ki}
De-Chang Dai, Cigdem Issever, Eram Rizvi, Glenn Starkman, Dejan Stojkovic, and
  Jeff Tseng.
\newblock Black{M}ax: {A} black-hole event generator with rotation, recoil,
  split branes and brane tension.
\newblock {\em Phys. Rev.}, D77:076007, 2008.

\bibitem{Choptuik:2009ww}
Matthew~W. Choptuik and Frans Pretorius.
\newblock Ultra relativistic particle collisions.
\newblock {\em Phys. Rev. Lett.}, 104:111101, 2010.

\bibitem{Gingrich:2009hj}
Douglas~M. Gingrich.
\newblock Quantum black holes with charge, colour, and spin at the {LHC}.
\newblock {\em J. Phys.}, G37:105108, 2010.

\bibitem{Witek:2010xi}
Helvi Witek, Miguel Zilhao, Leonardo Gualtieri, Vitor Cardoso, Carlos Herdeiro,
  Andrea Nerozzi, and Ulrich Sperhake.
\newblock Numerical relativity for {D} dimensional space-times: head- on
  collisions of black holes and gravitational wave extraction.
\newblock {\em Phys. Rev.}, D82:104014, 2010.

\bibitem{Witek:2010az}
Helvi Witek, Vitor Cardoso, Leonardo Gualtieri, Carlos Herdeiro, Ulrich
  Sperhake, and Miguel Zilhao.
\newblock Head-on collisions of unequal mass black holes in {D}=5 dimensions.
\newblock {\em Phys. Rev.}, D83:044017, 2011.

\bibitem{Nayeri:2005ck}
Ali Nayeri, Robert~H. Brandenberger, and Cumrun Vafa.
\newblock Producing a scale-invariant spectrum of perturbations in a {H}agedorn
  phase of string cosmology.
\newblock {\em Phys. Rev. Lett.}, 97:021302, 2006.

\bibitem{Brandenberger:2011et}
Robert~H. Brandenberger.
\newblock String gas cosmology: {P}rogress and problems.
\newblock arXiv:1105.3247, 2011.

\bibitem{ArkaniHamed:1998nn}
Nima Arkani-Hamed, Savas Dimopoulos, and G.~R. Dvali.
\newblock Phenomenology, astrophysics and cosmology of theories with
  sub-millimeter dimensions and {T}e{V} scale quantum gravity.
\newblock {\em Phys. Rev.}, D59:086004, 1999.

\end{thebibliography}

\appendix


\section{Proof that extrinsic curvatures are tetrad-frame tensors}
\label{proofextrinsic}

The tetrad frame is fixed so that the outgoing and ingoing null axes
$\bgamma_v$ and $\bgamma_u$
point along the special outgoing and ingoing null geodesic directions.
The remaining unfixed tetrad gauge transformations
are a Lorentz boost in the $vu$ plane, which transforms
\begin{equation}
\label{tetradvutrans}
  \bgamma_v \rightarrow \ee^\nu \bgamma_v
  \ , \quad
  \bgamma_u \rightarrow \ee^{- \nu} \bgamma_u
  \ ,
\end{equation}
and a spatial rotation in the $+-$ plane, which transforms
\begin{equation}
\label{tetradpmtrans}
  \bgamma_+ \rightarrow \ee^{\im \mu} \bgamma_+
  \ , \quad
  \bgamma_- \rightarrow \ee^{- \im \mu} \bgamma_-
  \ .
\end{equation}
Tetrad-frame connection coefficients
are defined by
$\Gamma_{klm} \equiv \bgamma_k \cdot \partial_m \bgamma_l$.
In particular, the extrinsic curvature component
$\Gamma_{+v-}$ (for example)
is defined by
\begin{equation}
\label{Gammapvmdef}
  \Gamma_{+v-}
  \equiv
  \bgamma_+ \cdot \partial_- \bgamma_v
  \ .
\end{equation}
Upon a tetrad transformation~(\ref{tetradvutrans}) and (\ref{tetradpmtrans}),
the only potentially non-tensorial contribution to the transformation
of the extrinsic curvature component~(\ref{Gammapvmdef}) is
\begin{equation}
  \bgamma_+ \cdot \bgamma_v \, \partial_- \ee^{\nu}
  \ ,
\end{equation}
but this vanishes because the angular and null axes
$\bgamma_+$ and $\bgamma_v$
are orthogonal.
A similar argument applies to each of the
16 components of the extrinsic curvatures in each of the 4 tetrad directions.
Thus the extrinsic curvatures form tetrad-frame tensors as claimed.

\section{Accuracy of the Hamilton-Jacobi approximation for geodesics}
\label{accuracyHJ}

This Appendix confirms the accuracy of the
the Hamilton-Jacobi approximation~(\ref{pk}), (\ref{Pk})
for particles moving in the spacetime defined
by the line-element~(\ref{lineelement}).
For simplicity and brevity,
the demonstration is restricted to massless particles.
It is clear physically that,
under the hyper-relativistic conditions characteristic of inflation,
the motions of massive particles should be well-approximated
by those of massless particles,
as is demonstrated explictly by \cite{Hamilton:2010b}
for the case of conformally separable spacetimes.

For a massless particle, $m = 0$,
the Hamilton-Jacobi approximation~(\ref{pk}), (\ref{Pk})
to the tetrad-frame momentum $p_k$ of a freely-falling particle
yields the following expressions for the
covariant derivatives $\DD p_k / \DD \lambda$
with respect to affine parameter $\lambda$ along the trajectory:
\begin{subequations}
\label{DpHJ}
\begin{align}
  p^k {\DD p_k \over \DD \lambda}
  &=
  0
  \ ,
\\
\label{DpkHJ}
  {\DD p_k \over \DD \lambda}
  &=
  {\KCarter \over \rhoy^2}
  \,
  \partial_k \ln \left( {\rhox \over \rhoy} \right)
  \quad
  ( k = t , y , \phi )
  \ .
\end{align}
\end{subequations}
The covariant derivatives would vanish if the path were exactly geodesic;
they do not vanish because the Hamilton-Jacobi approximation is not exact.
The covariant derivatives do vanish along the principal null directions,
which have vanishing Carter constant, $\KCarter = 0$.
The covariant derivatives would also vanish if the two conformal factors
were equal, $\rhox = \rhoy$, which is the conformally separable case.

The Hamilton-Jacobi approximation~(\ref{pk}), (\ref{Pk})
can be considered adequate if,
integrated along the path of a particle,
the difference $\Delta p_k$ between the approximate
Hamilton-Jacobi momentum $p_k$, which satisfies equations~(\ref{DpHJ}),
and the true momentum, which satisfies $\DD p_k / \DD \lambda = 0$,
is small,
$\Delta p_k \ll p_k$.
Appendix~D of \cite{Hamilton:2010b}
gives general criteria by which such integrals may be judged small.
The integrals of equations~(\ref{DpkHJ}) easily satisfy those
smallness criteria
(specifically,
the parameters $\alpha$ and $\beta$ in Appendix~D of \cite{Hamilton:2010b}
are
$\alpha = 0$, $\beta = -1$ for $p_y$ and $p_\phi$,
and
$\alpha = 1/2$, $\beta = -5/2$ for $p_t$).

The Hamilton-Jacobi approximation to the number density $N$
along a collisionless stream is given by equation~(\ref{NHJ}).
The corresponding number current $n_k$, equation~(\ref{nk}),
satisfies
\begin{align}
  {1 \over N}
  \DD^k n_k
  &=
  -
  \left[
  {P_x^2 - P_t^2 \over P_x^2}
  p^t
  \partial_t
  \right.
\nonumber
\\
\label{DknkHJ}
  &\quad
  \left.
  + \,
  {P_x^2 + P_t^2 \over P_x^2}
  \left(
  p^y
  \partial_y
  +
  p^\phi
  \partial_\phi
  \right)
  \right]
  \ln \left( {\rhox \over \rhoy} \right)
  \ .
\end{align}
The true number current should be covariantly conserved,
$D^k n_k = 0$;
the Hamilton-Jacobi number current is not conserved,
equation~(\ref{DknkHJ}),
because the Hamilton-Jacobi approximation is not exact.
Since
\begin{equation}
  {\dd \ln N \over \dd \lambda}
  =
  {1 \over N}
  D_k n^k
  -
  D_k p^k
  \ ,
\end{equation}
and the momentum $p^k$ has already been checked
to be given accurately by
the Hamilton-Jacobi approximation~(\ref{pk}), (\ref{Pk}),
the Hamilton-Jacobi approximation~(\ref{NHJ}) to the number density $N$
can be deemed adequate if the right hand side of equation~(\ref{DknkHJ}),
integrated over the path of a particle, is small.
Again,
the integral easily satisfies the smallness criteria given in
Appendix~D of \cite{Hamilton:2010b}
(the $p^t$ term on the right hand side has $\alpha = 1$, $\beta = 0$,
while the $p^y$ and $p^\phi$ terms have $\alpha = 1/2$, $\beta = -3/2$).

\end{document}